\documentclass[aps,pre,showpacs,floatfix]{revtex4}
\usepackage{epsf}
\usepackage{subfigure}
\usepackage{amsmath}
\usepackage{amssymb}
\usepackage{bm}
\usepackage{graphicx}
\usepackage{epstopdf}
\newcommand{\mean}[1]{\left \langle #1 \right \rangle}

\begin{document}

\newcommand{\Q}{{\widetilde Q}}

\newcommand{\be}{\begin{equation}}
\newcommand{\ee}{\end{equation}}
\newcommand{\bea}{\begin{eqnarray}}
\newcommand{\eea}{\end{eqnarray}}
\newcommand{\cum}[1]{\left \langle \left \langle #1 \right \rangle 
\right \rangle}

\renewcommand{\d}{\text{d}}
\newcommand{\e}{\text{e}}
\renewcommand{\i}{\text{i}}
\newcommand{\rint}{\int\displaylimits}
\renewcommand{\Re}{\mathop{\text{Re}}\nolimits}
\newcommand{\tr}{\mathop{\text{tr}}\nolimits}
\newcommand{\Tr}{\mathop{\text{Tr}}\nolimits}
\newcommand{\ket}[1]{|{#1}\rangle}
\newcommand{\bra}[1]{\langle{#1}|}
\newcommand{\bras}[2]{{}_{#2}\hspace*{-0.2mm}\langle{#1}|}
\newcommand{\bracket}[2]{\langle#1|#2\rangle}
\newcommand{\Det}{\mathop{\text{Det}}\nolimits}
\newcommand{\pv}{\mathop{\text{P}}\nolimits}
\newcommand{\erf}{\mathop{\text{erf}}\nolimits}
\newcommand{\erfc}{\mathop{\text{erfc}}\nolimits}
\newcommand{\erfi}{\mathop{\text{erfi}}\nolimits}
\newcommand{\sinc}{\mathop{\text{sinc}}\nolimits}

\renewcommand{\labelenumi}{(\roman{enumi})}
\renewcommand{\labelitemii}{$\triangleright$}

\newcommand{\cmt}[1]{\textbf{[#1]}}

\title{\bf Fluctuation theorem for currents in open quantum systems}

\author{David Andrieux and Pierre Gaspard}
\affiliation{Center for Nonlinear Phenomena and Complex Systems,\\
Universit\'e Libre de Bruxelles, Code Postal 231, Campus Plaine,
B-1050 Brussels, Belgium \\
{\rm E-mail: dandrieu@ulb.ac.be; gaspard@ulb.ac.be}}

\author{Takaaki Monnai and Shuichi Tasaki}
\affiliation{Department of Applied Physics,\\
Waseda University, 3-4-1 Okubo, Shinjuku-ku, Tokyo 169-8555, Japan \\
{\rm E-mail: t.monnai@kurenai.waseda.jp; stasaki@waseda.jp}}

\begin{abstract}
A quantum-mechanical framework is set up to describe the full counting
statistics of particles flowing between reservoirs in an open system
under time-dependent driving.  A symmetry relation is obtained which is
the consequence of microreversibility for the probability of the nonequilibrium
work and the transfer of particles and energy between the reservoirs.
In some appropriate long-time limit, the symmetry relation leads to a 
steady-state
quantum fluctuation theorem for the currents between the reservoirs.
On this basis, relationships are deduced which extend the Onsager-Casimir
reciprocity relations to the nonlinear response coefficients.
\end{abstract}



\maketitle

\section{Introduction}

Quantum systems can be driven out of equilibrium by time-dependent
perturbations, by interaction with reservoirs at different chemical 
potentials or temperatures, or by combining both.
In the latter cases, the quantum system is open and currents of energy 
and particles
are flowing across the system. Such processes take place
in mesoscopic electronic conductors as well as in chemical reactions.
These nonequilibrium processes can be characterized by the relations
linking their currents to the differences of chemical potentials. 
These relations may be linear
in the case of Ohm's law, but are typically nonlinear, which defines 
the nonlinear response
coefficients.

Alternatively, the full counting statistics of the particles 
transferred between
the reservoirs can be considered.  This statistics aims to 
characterize the transfers of particles
in terms of the functions generating all the statistical moments of 
the fluctuating numbers
of particles. The knowledge of this generating function gives access 
not only to the conductance
and the noise power but also to higher-order moments and thus to the 
properties of nonlinear response.
The full counting statistics has attracted considerable theoretical 
interest and is also envisaged in experimental measurements 
\cite{OKLY99,FHTH06,GLSSISEDG06}.
After the pioneering work of Ref. \cite{LL93}, several methods have 
been developed in order to obtain the full counting statistics in 
mesoscopic conductors.
One of them is based
on Keldysh Green's function formalism, in which an expression for the 
generating function
has been obtained within a semiclassical approximation 
\cite{NB02,K03,N99,BN03,N07}.
The full counting statistics can also be obtained on the basis of quantum
Markovian master equations describing the fluctuations of the 
currents \cite{N02}, as well as
in terms of stochastic path integrals \cite{JSP04}.
The generating function obtained in the approaches using the 
semiclassical approximation or
the Markovian master equation has been shown to obey a symmetry 
relation as the consequence
of time reversibility \cite{TN05}.  
In nonequilibrium statistical mechanics, 
this relation is known as the fluctuation theorem which 
has been established for several classes of systems.  These latter are either
time-independent deterministic \cite{ECM93,GC95,G96} or stochastic 
systems sustaining nonequilibrium steady states 
\cite{K98,LS99,M99,MN03,CVK06,AG04,AG06JSM,AG07JSP},
or time-dependent Hamiltonian or stochastic systems, in which case 
the fluctuation theorem is closely
related to Jarzynski nonequilibrium work theorem \cite{J97,C99,KPV07}.
Similar symmetry relations have been considered for continuous-time random walks \cite{EL08}.  
Quantum versions of the fluctuation theorem
and Jarzynski nonequilibrium work theorem have also been obtained 
\cite{K00,T00,M05,TLH07,TH07,MT03,M03,DM04,EM06,HEM06,EHM07E,EHM07B,EHM08,CV07,SD07,SU08}. 
Moreover, a further relationship has recently been proved
for time-dependent quantum Hamiltonian systems \cite{AG08},
allowing the derivation of Green-Kubo formulas 
and Onsager-Casimir reciprocity relations for
the linear response coefficients \cite{G52,K57,O31,C45}.

An open question is to bridge the gap separating the time-dependent situations 
from the nonequilibrium steady states which are expected to be reached in the 
long-time limit.
The problem is to deal with nonequilibrium steady states without 
relying on the semiclassical or Markovian approximations, or on the 
neglect of the energy or particle content of the subsystem
coupling the reservoirs.

In the present paper, our purpose is to prove the fluctuation theorem 
for the currents in open systems obeying Hamiltonian quantum dynamics 
and sustaining nonequilibrium steady states in the long-time limit. 
We start by considering a time-dependent quantum system in contact with energy 
and particle reservoirs at
different temperatures and chemical potentials.  The amounts of 
energy and particles
which are exchanged between the initial and final times are determined by
quantum measurements. This framework is similar to the one considered
by Kurchan to obtain a fluctuation theorem for quantum systems \cite{K00}. 
Here, this framework is extended by taking the initial 
states as grand-canonical instead of canonical equilibrium states, 
which allows us to deal with transfers of particles between the reservoirs.
In this way, we obtain an exact relationship which is the consequence 
of microreversibility
for the probability of a certain exchange of energy and particles 
between the reservoirs
during the time-dependent external drive. An equivalent symmetry 
relation is obtained for the generating function of all the 
fluctuating variables. However, these symmetry relations are 
expressed in terms of the temperatures and chemical potentials of the 
reservoirs.  The problem is that we need a symmetry relation in terms 
of the {\it differences} of temperatures and chemical potentials, 
which define the thermodynamic forces (also called the affinities) driving 
the currents across the system. The importance of this point 
has recently been discussed in the review \cite{EHM08}.

The central contribution of the present paper is the proof that, in the 
long-time limit, the aforementioned generating function only depends 
on the  {\it differences} between the parameters of the reservoirs. 
This proof is carried out by obtaining lower 
and upper bounds on the generating function in terms of a new 
generating function which only depends on the {\it differences} of 
parameters and further functions which are bounded in the long-time limit.
Combining this fundamental result with the previously established 
symmetry relation of the generating function, the fluctuation theorem 
is proved for nonequilibrium steady states in open quantum systems.
Thanks to this quantum fluctuation theorem, the Onsager-Casimir 
reciprocity relations and their generalizations to the nonlinear 
response coefficients can be inferred \cite{AG04,AG07JSM}.

The plan of the paper is the following.  The protocols for the 
forward and reversed drives of the open system are introduced in Sec. 
\ref{protocols}.  The symmetry relations for the probability
and the generating function are proved in Sec. \ref{symmetry}.  In 
Sec. \ref{QFT}, we obtain
the quantum fluctuation theorem for the currents in the steady state 
reached in the long-time limit.
In Sec. \ref{NLR}, the consequences of the fluctuation theorem
on the linear and nonlinear response coefficients are deduced.
The conclusions are drawn in Sec. \ref{Conclusions}.

\section{Open quantum system and time-dependent protocols}
\label{protocols}

\subsection{The total Hamiltonian}

We consider a total quantum system composed of a subsystem in contact 
with several reservoirs of energy and particles.  Initially, the 
reservoirs are decoupled. During a time interval $\cal T$, the 
reservoirs are put in contact by some time-dependent interaction 
which has the effect of changing the energy and the particle numbers 
in each reservoir.  The total Hamiltonian of the system is thus given 
by
\bea
H(t;B) &= & \sum_{j=1}^r H_j \qquad \mbox{for} \quad t < 0 \; ,\\
&= & \sum_{j=1}^r H_j + V(t) \qquad \mbox{for} \quad 0 \leq t < {\cal T} \; ,\\
&= & \sum_{j=1}^r \tilde H_j \qquad \mbox{for} \quad {\cal T} \leq t \; ,
\eea
where $H_j$ denotes the Hamiltonian of the $j^{\rm th}$ isolated 
reservoir before the interaction is switched on.  During the time 
interval $0\leq t <{\cal T}$, the system is submitted to the 
time-dependent interaction $V(t)$, which describes the coupling of 
the reservoirs by the subsystem.  Beyond the final time $t={\cal T}$, 
the reservoirs are decoupled into the Hamiltonians $\tilde H_j$.  We 
suppose that the whole system is placed in a magnetic field $B$.

The observables of the reservoirs are the Hamiltonian operators $H_j$ 
and $\tilde H_j$, as well as the numbers of particles of several 
species $N_{j\alpha}$ with $\alpha=1,2,...,c$.
The total number of particles of species $\alpha$ is given by
\be
N_{\alpha} = n_{\alpha} + \sum_{j=1}^r N_{j \alpha} \; ,
\ee
where $n_{\alpha}$ denotes the number of particles of species 
$\alpha$ in the subsystem between the reservoirs.

Since the numbers of particles of species $\alpha$ are conserved 
within each isolated reservoirs before and after their coupling, we 
have that
\be
\left[ H_j, N_{j'\alpha} \right] = 0 \qquad \mbox{and}\qquad \left[ 
\tilde H_j, N_{j'\alpha} \right] = 0 \; ,
\ee
for all $j,j'=1,2,...,r$ and $\alpha=1,2,...,c$.

We suppose that the Hamiltonian operator $H(t;B)$ has the symmetry
\bea
\Theta H(t;B) \Theta = H(t; -B)
\label{TR}
\eea
under the time-reversal operator $\Theta$.  This latter is an antilinear
operator such that $\Theta^2=I$ and which has the effect of changing the
sign of all odd parameters such as magnetic fields.
Equation (\ref{TR}) expresses the microreversibility in an external 
magnetic field.
The numbers of particles are symmetric under time reversal:
\bea
\Theta N_{j\alpha} \Theta = N_{j\alpha} \qquad \mbox{and}\qquad 
\Theta n_{\alpha} \Theta = n_{\alpha}\, .
\eea

In the following, we consider protocols with two quantum measurements
at the initial and final times separated by a unitary time evolution
(see Ref. \cite{EHM08} for a review).

\subsection{The forward protocol}

The forward time evolution is defined as
\bea
i\hbar \frac{\partial}{\partial t} U_{\rm F} (t;B) = H(t;B)U_{\rm F} (t;B) \; ,
\label{U_F}
\eea
with the initial condition $U_{\rm F} (0;B)=I$.
In the Heisenberg representation, the observables evolve according to
\bea
A_{\rm F}(t)=U^{\dagger}_{\rm F}(t;B) \, A\,  U_{\rm F}(t;B) \; ,
\eea
which also concerns the time-dependent Hamiltonian
\bea
H_{\rm F}(t)=U^{\dagger}_{\rm F}(t;B) H(t;B) U_{\rm F}(t;B) \, .
\eea
The average of an observable is thus obtained from
\bea
\mean{A_{\rm F}(t)}= {\rm tr} \rho(0;B) A_{\rm F}(t) \, .
\eea
We note that the dependence on the magnetic field is implicit in 
these expressions.

The initial state of the system is taken as the following 
grand-canonical equilibrium state of the decoupled reservoirs at the 
different inverse temperatures $\beta_j=1/(k_{\rm B}T_j)$ and 
chemical potentials $\mu_{j\alpha}$:
\be
\rho(0;B) = \prod_{j=1}^r \frac{{\rm 
e}^{-\beta_j(H_j-\sum_{\alpha}\mu_{j\alpha} N_{j\alpha})}}{\Xi_j(B)} 
= \prod_{j=1}^r {\rm e}^{-\beta_j\left[H_j-\sum_{\alpha}\mu_{j\alpha} 
N_{j\alpha}-\Phi_j(B)\right]}
\label{rho_0} \; ,
\ee
where $\Phi_j(B)= -k_{\rm B}T_j \ln \Xi_j(B)$ denotes the 
thermodynamic grand-potential of the $j^{\rm th}$ reservoir in the 
initial equilibrium state.

A quantum measurement is performed at the initial time.  The system 
is observed in the eigenstate $\vert\Psi_k\rangle$ of the reservoir 
operators of energy and particle numbers:
\bea
t=0: \qquad  H_j \vert\Psi_k\rangle &=& \epsilon_{jk} \vert\Psi_k\rangle \; ,\\
N_{j\alpha} \vert\Psi_k\rangle &=& \nu_{j\alpha k} \vert\Psi_k\rangle \; .
\eea

At the final time, another quantum measurement is performed and the 
system is observed in the eigenstate $\vert\tilde\Psi_l\rangle$ of 
the reservoir operators of energy and particle numbers:
\bea
t={\cal T}: \qquad  \tilde H_j \vert\tilde\Psi_l\rangle &=& 
\tilde\epsilon_{jl} \vert\tilde\Psi_l\rangle \; ,\\
N_{j\alpha} \vert\tilde\Psi_l\rangle &=& \tilde\nu_{j\alpha l} 
\vert\tilde\Psi_l\rangle \; .
\eea

Accordingly, during the forward time evolution, the following change 
of energy in the $j^{\rm th}$ reservoir is observed:
\be
\Delta \epsilon_j = \tilde\epsilon_{jl} - \epsilon_{jk} \; ,
\label{De}
\ee
while the number of particles of species $\alpha$ in the $j^{\rm th}$ 
reservoir changes by
\be
\Delta \nu_{j\alpha} = \tilde \nu_{j\alpha l} - \nu_{j\alpha k} \; .
\label{Dn}
\ee

\subsection{The reversed protocol}

The evolution operator of the reversed process is defined as
\bea
i\hbar \frac{\partial}{\partial t} U_{\rm R} (t;B) = H({\cal 
T}-t;B)U_{\rm R} (t;B) \, ,
\label{UR}
\eea
with the initial condition $U_{\rm R} (0;B)=I$,
and is related to the one of the forward process by the following
\\

{\bf Lemma:}
{\it The forward and reversed time evolution operators at the final 
time $\cal T$ are related to each other by}
\bea
U_{\rm R} ({\cal T};-B) = \Theta U^{\dagger}_{\rm F} ({\cal T};B) \Theta \, .
\label{R-F}
\eea
\\

This lemma is proved by noting that the forward time evolution in the 
magnetic field $B$, followed by the operation of time reversal, by 
the reversed time evolution in the magnetic field $-B$, and finally 
by time reversal again is equal to the identical operator:
\bea
\Theta U_{\rm R} ({\cal T};-B) \Theta U_{\rm F} ({\cal T};B) = I \; ,
\eea
from which we deduce Eq. (\ref{R-F}).\\

The reversed protocol is supposed to start with the following 
grand-canonical equilibrium state of the final decoupled reservoirs:
\be
\rho({\cal T};-B) = \prod_{j=1}^r \frac{{\rm e}^{-\beta_j(\tilde 
H_j-\sum_{\alpha}\mu_{j\alpha} N_{j\alpha})}}{\tilde\Xi_j(-B)} = 
\prod_{j=1}^r {\rm e}^{-\beta_j\left[\tilde 
H_j-\sum_{\alpha}\mu_{j\alpha} N_{j\alpha}-\tilde\Phi_j(-B)\right]} 
\; ,
\label{rho_T}
\ee
at the same inverse temperatures $\beta_j=1/(k_{\rm B}T_j)$ and 
chemical potentials $\mu_{j\alpha}$ as in the forward protocol and 
where $\tilde\Phi_j(-B)= -k_{\rm B}T_j \ln \tilde\Xi_j(-B)$ denotes 
the grand-canonical thermodynamic potential of the $j^{\rm th}$ 
reservoir in the final equilibrium state and reversed magnetic field.

Similarly to the forward protocol, quantum measurements are performed
at the initial and final times to determine the changes of energies 
and particle numbers in the reservoirs.

\section{Consequences of microreversibility}
\label{symmetry}

\subsection{The symmetry relation for the probability of the fluctuations}

The probability distribution function to observe the energy 
(\ref{De}) and particle transfers (\ref{Dn}) during the forward 
protocol is defined as
\bea
p_{\rm F} (\Delta \epsilon_j,\Delta \nu_{j\alpha};B) &\equiv& 
\sum_{kl} \; \prod_j \delta\left[\Delta\epsilon_j - 
(\tilde\epsilon_{jl} - \epsilon_{jk})\right] \; \prod_{j\alpha} 
\delta\left[ \Delta \nu_{j\alpha} - (\tilde \nu_{j\alpha l} - 
\nu_{j\alpha k})\right] \nonumber\\ && \times \vert \langle 
\tilde\Psi_l(B) \vert U_{\rm F}(T;B)\vert\Psi_k(B)\rangle\vert^2 \; 
\langle \Psi_k(B)\vert \rho(0;B)\vert\Psi_k(B)\rangle \; .
\eea
We notice that this function is a probability density because the 
quantities $\delta(\cdot)$ are Dirac distributions for both the 
energy and the particle numbers.

Inserting the expression of the initial density matrix (\ref{rho_0}), 
using the Dirac delta distributions to replace the initial energies 
and numbers into the final ones, we find that
\bea
p_{\rm F} (\Delta \epsilon_j,\Delta n_{\alpha};B) &=& \sum_{kl} \; 
\prod_j \delta\left[\Delta\epsilon_j - (\tilde\epsilon_{jl} - 
\epsilon_{jk})\right] \; \prod_{j\alpha} \delta\left[ \Delta 
\nu_{j\alpha} - (\tilde \nu_{j\alpha l} - \nu_{j\alpha k})\right] 
\nonumber \\ && \times
\vert \langle \tilde\Psi_l(B) \vert U_{\rm F}(T;B)\vert\Psi_k(B)\rangle\vert^2
\; {\rm e}^{-\sum_j\beta_j\left[\epsilon_{jk}-\sum_{\alpha} 
\mu_{j\alpha}\nu_{j\alpha k} -\Phi_j(B)\right]}
\nonumber\\
\nonumber\\
&=&
{\rm 
e}^{\sum_j\beta_j(\Delta\epsilon_{j}-\sum_{\alpha}\mu_{j\alpha}\Delta 
\nu_{j\alpha}-\Delta\Phi_{j})}
\nonumber \\ && \times
\sum_{kl} \; \prod_j \delta\left[\Delta\epsilon_j - 
(\tilde\epsilon_{jl} - \epsilon_{jk})\right] \; \prod_{j\alpha} 
\delta\left[ \Delta \nu_{j\alpha} - (\tilde \nu_{j\alpha l} - 
\nu_{j\alpha k})\right]
\nonumber \\ && \times
\vert \langle \tilde\Psi_l(B)\vert U_{\rm F}(T;B)\vert\Psi_k(B)\rangle\vert^2
\,  {\rm e}^{-\sum_j\beta_j\left[\tilde\epsilon_{jl}-\sum_{\alpha} 
\mu_{j\alpha}\tilde \nu_{j\alpha l} -\tilde\Phi_j(-B)\right]} \; ,
\nonumber\\
\label{Id1}
\eea
where we have introduced the difference of the thermodynamic 
grand-potential of the $j^{\rm th}$ reservoir as
\be
\Delta\Phi_{j} \equiv \tilde \Phi_j(-B)- \Phi_j(B) \; .
\ee
According to the lemma (\ref{R-F}), the probability of the transition 
$k\to l$ during the forward process is equal to the probability of 
the transition $l\to k$ in the reversed process and magnetic field:
\bea
\vert \langle \tilde\Psi_l(B)\vert U_{\rm F}({\cal 
T};B)\vert\Psi_k(B)\rangle\vert^2 &=&
\vert \langle \tilde\Psi_l(B)\vert \Theta U_{\rm R}^{\dagger}({\cal 
T};-B)\Theta\vert\Psi_k(B)\rangle\vert^2 \nonumber \\ &=&
\vert \langle \Psi_k(-B)\vert U_{\rm R}({\cal 
T};-B)\vert\tilde\Psi_l(-B)\rangle\vert^2  \; .
\eea
Substituting this identity in Eq. (\ref{Id1}) and introducing the 
probability of negative changes in the energies and particle numbers 
during the reversed process as
\bea
p_{\rm R} (-\Delta \epsilon_j,-\Delta \nu_{j\alpha};-B) &\equiv& 
\sum_{kl} \; \prod_j \delta\left[-\Delta\epsilon_j - 
(\epsilon_{jk}-\tilde\epsilon_{jl})\right] \; \prod_{j\alpha} 
\delta\left[ -\Delta \nu_{j\alpha} - (\nu_{j\alpha k}-\tilde 
\nu_{j\alpha l})\right]
\nonumber \\ && \times
\vert \langle \Psi_k(-B)\vert U_{\rm R}({\cal 
T};-B)\vert\tilde\Psi_l(-B)\rangle\vert^2
\, \langle \tilde\Psi_l(-B)\vert \rho({\cal T};-B)\vert\tilde\Psi_l(-B)\rangle \nonumber\\
\label{p_R}
\eea
with the final density matrix (\ref{rho_T}), we obtain the following 
symmetry relation:
\bea
p_{\rm F} (\Delta \epsilon_j,\Delta \nu_{j\alpha};B) ={\rm 
e}^{\sum_j\beta_j(\Delta\epsilon_{j}-\sum_{\alpha}\mu_{j\alpha}\Delta\nu_{j\alpha}-\Delta\Phi_{j})} 
\,
p_{\rm R} (-\Delta \epsilon_j,-\Delta \nu_{j\alpha};-B) \; .
\label{PFT}
\eea
If this relation is restricted to the energy change in a single 
system, this fluctuating quantity is the work $W$ performed on the 
system and we recover the quantum version of Crooks' fluctuation 
theorem
\bea
p_{\rm F} (W;B) ={\rm e}^{\beta(W-\Delta F)} \, p_{\rm R} (-W;-B)
\label{CrooksFT}
\eea
with the corresponding difference of free energy $\Delta F=\tilde 
F(-B)-F(B)$ \cite{C99}.
The relation (\ref{PFT}) extends this result to the transfer of 
particles under the effect of
the differences of chemical potentials driving the system out of equilibrium.

\subsection{The symmetry relation for the generating function}

The generating functions of the statistical moments of the exchanges 
of energy and particles are defined by
\be
G_{\rm F,R}(\xi_j,\eta_{j\alpha};B) \equiv \int \prod_{j\alpha} 
d\Delta\epsilon_j \, d\Delta \nu_{j\alpha} \, {\rm e}^{-\sum_j\xi_j 
\Delta\epsilon_j -\sum_{\alpha} \eta_{j\alpha}\Delta \nu_{j\alpha}} \,
p_{\rm F,R} (\Delta \epsilon_j,\Delta \nu_{j\alpha};B)
\ee
for the forward and reversed processes. The knowledge of these 
generating functions provides the full counting statistics of the 
process.
We notice that the generating function of the forward protocol is 
alternatively defined as
\be
G_{\rm F}(\xi_j,\eta_{j\alpha};B) = \left\langle {\rm 
e}^{-\sum_j\xi_j \tilde H_{j{\rm F}} - \sum_{j\alpha} 
\eta_{j\alpha}N_{j\alpha{\rm F}}} {\rm e}^{\sum_j\xi_j H_{j} + 
\sum_{j\alpha} \eta_{j\alpha}N_{j\alpha}}\right\rangle_{\rm F}
\ee
with
\bea
\tilde H_{j{\rm F}} &=& U_{\rm F}^{\dagger}({\cal T};B) \tilde H_{j} 
U_{\rm F}({\cal T};B) \label{HF} \; ,\\
N_{j\alpha{\rm F}} &=& U_{\rm F}^{\dagger}({\cal T};B) 
N_{j\alpha}U_{\rm F}({\cal T};B) \label{NaF} \; ,
\eea
and the generating function of the reversed protocol as
\be
G_{\rm R}(\xi_j,\eta_{j\alpha};-B) = \left\langle {\rm 
e}^{-\sum_j\xi_j H_{j{\rm R}} - \sum_{j\alpha} 
\eta_{j\alpha}N_{j\alpha{\rm R}}} {\rm e}^{\sum_j\xi_j \tilde H_{j} + 
\sum_{j\alpha} \eta_{j\alpha}N_{j\alpha}}\right\rangle_{\rm R}
\ee
with
\bea
H_{j{\rm R}} &=& U_{\rm R}^{\dagger}({\cal T};-B) H_{j} U_{\rm 
R}({\cal T};-B) \label{HR} \; , \\
N_{j\alpha{\rm R}} &=& U_{\rm R}^{\dagger}({\cal T};-B) 
N_{j\alpha}U_{\rm R}({\cal T};-B) \label{NaR} \; .
\eea

Taking the Laplace transforms of the symmetry relation (\ref{PFT}), 
we obtain an equivalent symmetry relation in terms of the generating 
functions:
\be
G_{\rm F}(\xi_j,\eta_{j\alpha};B) = {\rm 
e}^{-\sum_j\beta_j\Delta\Phi_{j}} \, G_{\rm 
R}(\beta_j-\xi_j,-\beta_j\mu_{j\alpha}-\eta_{j\alpha};-B) \; ,
\label{GFT}
\ee
in terms of the temperatures and chemical potentials of the reservoirs.
As mentioned in the introduction, this symmetry relation has not yet 
the appropriate form
because the thermodynamic forces or affinities do not appear.

\section{Quantum fluctuation theorem for the currents}
\label{QFT}

In this section, we prove that, in the long-time limit, the 
generating functions entering into
the symmetry relation (\ref{GFT}) only depend on the differences of 
the parameters $\xi_j$ and $\eta_{j\alpha}$, leading to the requested 
symmetry.
In the long-time limit, a nonequilibrium steady state can be reached 
between the reservoirs
if the coupling remains constant over the whole time interval except 
finite transients.

\subsection{The theorem}

We consider a situation where two large quantum systems interact through a
bounded time-dependent perturbation described by $V(t)$. Then, the generator of
time evolution of the whole system is given by
\begin{eqnarray}
H&=& H_1+H_2+V(t) \; .
\end{eqnarray}
Hereafter, we assume that $V(0)=0$, $V(t)=V({\cal T}-t)$ and $V(t)=V_0$ for
$t_0\le t\le {\cal T}-t_0$.

Let $N_{j\alpha}$ ($j=1,2$) be the number of $\alpha$-particles
in the $j^{\rm th}$ large system and assume that 
$[N_{1\alpha}+N_{2\alpha}+n_\alpha,
H_1+H_2+V_0]=0$ with a bounded $n_\alpha$ (imagine that a quantum dot 
is located
between the two electrodes).

Since the interaction is symmetric under time reversal $V(t)=V({\cal T}-t)$,
the evolution operator of the forward and reversed protocols are identical
\be
U_{\rm F}(t;B)=U_{\rm R}(t;B)\equiv U(t;B)
\label{U_R}
\ee
and therefore solutions of one and the same equation:
\be
i{\partial\over \partial t}U(t;B)=\left[ H_1+H_2+V(t)\right]U(t;B)
\ee
with the initial condition $U(0;B)=I$ and $\hbar=1$.
For the same reason, the initial and final reservoir Hamiltonians are 
the same, $H_j=\tilde H_j$ for all $j=1,2,...,r$, so that the initial 
and final density matrices have the same definition
\be
\rho(B) = \prod_{j=1}^r {\rm e}^{-\beta_j\left[H_j-\sum_{\alpha}\mu_{j\alpha} 
N_{j\alpha}-\Phi_j(B)\right]}\; ,
\label{rho}
\ee
where $\Phi_j(B)$ is the thermodynamic grand-potential of the $j^{\rm 
th}$ reservoir in magnetic field $B$. Accordingly, the forward and 
reversed generating functions also have the same definition
\begin{eqnarray}
G(\xi_j,\eta_{j\alpha};B)\equiv \left\langle {\rm 
e}^{-\sum_j\xi_jH_{j{\rm F}}-\sum_{j\alpha}\eta_{j\alpha}
N_{j\alpha{\rm F}}}
\ {\rm e}^{\sum_j\xi_jH_j+\sum_{j\alpha}\eta_{j\alpha}
N_{j\alpha}}\right\rangle
\end{eqnarray}
with Eqs. (\ref{HF}) and (\ref{NaF}) and where the average 
$\langle\cdot\rangle$ is carried out with respect to the density 
matrix (\ref{rho}).

According to Eq. (\ref{GFT}), this generating function has the symmetry
\be
G(\xi_j,\eta_{j\alpha};B) = {\rm e}^{-\sum_j\beta_j\Delta\Phi_{j}} \, 
G(\beta_j-\xi_j,-\beta_j\mu_{j\alpha}-\eta_{j\alpha};-B) \; ,
\label{GFT2}
\ee
in terms of the temperatures and chemical potentials of the reservoirs.

Our purpose in this section is to prove the

\medskip

\noindent{\bf Proposition.} {\it Assume that the limit
\be
Q(\xi_j,\eta_{j\alpha};B)\equiv -
\lim_{{\cal T}\to\infty}{1\over{\cal T}}\ln G(\xi_j,\eta_{j\alpha};B)
\label{Q-fn}
\ee
exists, it is a function only of $\xi_1-\xi_2$ and 
$\eta_{1\alpha}-\eta_{2\alpha}$:
\be
Q(\xi_j,\eta_{j\alpha};B)= {\widetilde 
Q}(\xi_1-\xi_2,\eta_{1\alpha}-\eta_{2\alpha};B) \; .
\label{dfn-tilde-Q}
\ee
}
\medskip

The interpretation of this proposition is that, 
because of the 
finiteness of the subsystem and the interaction $V_0$,
the energy and particles lost by the left (respectively right) 
reservoir 
are transferred to the right (respectively left) reservoir
within the overwhelming duration $t_0\le t \le {\cal T}-t_0$ and, as 
a result, $Q$ becomes a
function ${\widetilde Q}$ depending only on the differences 
$\xi_1-\xi_2$ and $\eta_{1\alpha}-\eta_{2\alpha}$.
We remark that the explicit form of the generating function 

${\widetilde Q}(\xi_1-\xi_2,\eta_{1\alpha}-\eta_{2\alpha};B)$
is given by Eq. (\ref{explicit}).

The above proposition implies that
\begin{eqnarray}
{\widetilde Q}(\xi_1-\xi_2,\eta_{1\alpha}-\eta_{2\alpha};B)
&=& Q(\xi_j,\eta_{j\alpha};B) \label{by-dfn-tilde-Q}\\
&=& Q(\beta_j-\xi_j, -\beta_j\mu_{j\alpha} -\eta_{j\alpha};-B)
\label{by-GFT2}\\
&=&{\widetilde 
Q}(\beta_1-\beta_2-\xi_1+\xi_2,-\beta_1\mu_{1\alpha}+\beta_2\mu_{2\alpha} 
-\eta_{1\alpha}+\eta_{2\alpha};-B)
\label{again-by-dfn-tilde-Q}\\
&=&{\widetilde Q}(A_0-\xi_1+\xi_2,A_\alpha
-\eta_{1\alpha}+\eta_{2\alpha};-B) \; ,
\label{final-FT}
\end{eqnarray}
where we have introduced the affinities:
\bea
A_0 &\equiv& \beta_1 - \beta_2 \; ,\label{A_0}\\
A_\alpha &\equiv& \beta_2\mu_{2\alpha}-\beta_1\mu_{1\alpha} \; ,\qquad 
\mbox{for} \quad \alpha=1,2,...,c,
\label{A_alpha}\eea
driving respectively the heat current and the
$\alpha$-particle currents from reservoir $2$ to reservoir $1$.
The result (\ref{final-FT}) is obtained by using the definition 
(\ref{dfn-tilde-Q}) at the line (\ref{by-dfn-tilde-Q}), the symmetry 
(\ref{GFT2}) and the independency of the quantities $\Delta\Phi_{j}$ 
on the time interval $\cal T$ at the line (\ref{by-GFT2}), again the 
definition (\ref{dfn-tilde-Q}) at the line 
(\ref{again-by-dfn-tilde-Q}), and finally the definitions of the 
affinities (\ref{A_0}) and (\ref{A_alpha}).
Hence, we have the

\medskip

\noindent{\bf Fluctuation theorem.} {\it The generating function of 
the independent currents satisfies the symmetry
\be
{\widetilde Q}(\xi,\eta_{\alpha};B) = {\widetilde 
Q}(A_0-\xi,A_{\alpha}-\eta_{\alpha};-B) \; .
\label{FT}
\ee
}
\medskip

In the particular case where the two systems have the same 
temperature, $\beta_1=\beta_2$,
the generating function has the symmetry:
\be
{\widetilde Q}(\xi,\eta_{\alpha};B)= {\widetilde Q}(-\xi, A_\alpha
-\eta_{\alpha};-B) \; ,
\ee
and we recover the symmetry
\be
{\widetilde Q}(0,\eta_{\alpha};B)= {\widetilde Q}(0, A_\alpha
-\eta_{\alpha};-B)
\ee
of the generating function of the independent particle currents, which has
already been proved elsewhere for stochastic processes \cite{AG07JSP}.

We notice that the fluctuation theorem (\ref{FT}) which is here proved thanks to
the proposition (\ref{Q-fn})-(\ref{dfn-tilde-Q}) reduces to the steady-state
fluctuation theorem presented as Eq. (104) in the review \cite{EHM08}
for vanishing magnetic field, $B=0$.
Accordingly, the proposition (\ref{Q-fn})-(\ref{dfn-tilde-Q})
also provides a rigorous proof of such steady-state fluctuation theorems.

\subsection{Setting}

In order to demonstrate the above proposition, the time evolution
is decomposed into different pieces corresponding to the short initial
transient over $0<t<t_0$, the long steady interaction over 
$t_0<t<{\cal T}-t_0$, and the final short transient over ${\cal 
T}-t_0<t<{\cal T}$. We introduce the lapse of time of the steady 
interaction
\be
\tau \equiv {\cal T}-2t_0 \; .
\ee

In addition to $U(t;B)$ defined by Eq. (\ref{U_F}), we introduce 
$U_1(t;B)$ as the solution of
\be
i{\partial\over \partial t}U_1(t;B)=\left[H_1+H_2+V(t_0-t)\right]U_1(t;B)
  \qquad \mbox{with}\quad U_1(0;B)=I \ .
\ee
It is then easy to show
\be
U(\tau+2t_0;B)=U_1(t_0;B){\rm e}^{-i{\bar H}\tau}U(t_0;B)\equiv U_f 
U_{\tau} U_i\ ,
\ee
where ${\bar H}=H_1+H_2+V_0$, $U_f=U_1(t_0;B)$, $U_{\tau}={\rm 
e}^{-i{\bar H}\tau}$
and $U_i=U(t_0;B)$.
We further note that
\begin{eqnarray}
&&{\rm e}^{iH_0(\tau+t_0)} U(\tau+2t_0;B){\rm e}^{iH_0t_0} = {\rm 
e}^{iH_0\tau} \Gamma_f{\rm e}^{-iH_0\tau} \Gamma_{\tau}{\rm 
e}^{-iH_0t_0}
\Gamma_i{\rm e}^{iH_0t_0} \ ,
\end{eqnarray}
where $H_0=H_1+H_2$, $\Gamma_f={\rm e}^{iH_0t_0}U_f$, 
$\Gamma_{\tau}={\rm e}^{iH_0\tau}U_{\tau}$, and $\Gamma_i={\rm 
e}^{iH_0t_0}U_i$.

Therefore, with the aid of $[H_0,H_j]=[H_0,N_{j\alpha}]=[H_0,\rho]=0$, we have
\begin{eqnarray}
&& G(\xi_j,\eta_{j\alpha};B) \nonumber\\ &=&
\langle U(\tau+2t_0;B)^\dag
{\rm e}^{-\sum_j\xi_jH_j-\sum_{j\alpha}\eta_{j\alpha}
N_{j\alpha}}\
U(\tau+2t_0;B)
\ {\rm e}^{\sum_j\xi_jH_j+\sum_{j\alpha}\eta_{j\alpha}
N_{j\alpha}}
\rangle
\nonumber\\
&=&
\langle U_i^\dag U_{\tau}^\dag U_f^\dag
{\rm e}^{-\sum_j\xi_jH_j-\sum_{j\alpha}\eta_{j\alpha}
N_{j\alpha}}\
U_f U_{\tau} U_i
\ {\rm e}^{\sum_j\xi_jH_j+\sum_{j\alpha}\eta_{j\alpha}
N_{j\alpha}}
\rangle
\nonumber\\
&=&
\langle {\rm e}^{-iH_0t_0}
U_i^\dag U_{\tau}^\dag U_f^\dag
{\rm e}^{-iH_0(\tau+t_0)}
{\rm e}^{-\sum_j\xi_jH_j-\sum_{j\alpha}\eta_{j\alpha}
N_{j\alpha}}\
{\rm e}^{iH_0(\tau+t_0)}
U_f U_{\tau} U_i
\ {\rm e}^{\sum_j\xi_jH_j+\sum_{j\alpha}\eta_{j\alpha}
N_{j\alpha}}
{\rm e}^{iH_0t_0}
\rangle
\nonumber\\
&=&
\langle \Gamma_i(-t_0)^\dag \Gamma_{\tau}^\dag \Gamma_f(\tau)^\dag
{\rm e}^{-\sum_j\xi_jH_j-\sum_{j\alpha}\eta_{j\alpha}
N_{j\alpha}}\
\Gamma_f(\tau) \Gamma_{\tau} \Gamma_i(-t_0)
\ {\rm e}^{\sum_j\xi_jH_j+\sum_{j\alpha}\eta_{j\alpha}
N_{j\alpha}} \rangle \; ,
\end{eqnarray}
where $\Gamma_\lambda(\tau)={\rm e}^{iH_0\tau} \Gamma_\lambda {\rm 
e}^{-iH_0\tau}$ ($\lambda=i$ or $f$).

For later purpose, we introduce
\begin{eqnarray}
&&\sum_j\xi_jH_j+\sum_{j\alpha}\eta_{j\alpha}N_{j\alpha}=2C+2D \; ,
\\
&&2C={\xi_1+\xi_2\over 2}H_0 +\sum_\alpha 
{\eta_{1\alpha}+\eta_{2\alpha}\over 2}N_{0\alpha} \; , \label{C}
\\
&&2D=(\xi_1-\xi_2)\Delta H_0 +\sum_\alpha 
(\eta_{1\alpha}-\eta_{2\alpha})\Delta N_{0\alpha} \; , \label{D}
\\
&&2A=\sum_j \beta_j\left(H_j-\sum_\alpha\mu_{j\alpha}N_{j\alpha}\right) \; ,
\end{eqnarray}
where $H_0=H_1+H_2$, $N_{0\alpha}=N_{1\alpha}+N_{2\alpha}$,
$\Delta H_0=(H_1-H_2)/2$, and $\Delta N_{0\alpha}=(N_{1\alpha}-N_{2\alpha})/2$.
Since $[C,\rho]=[D,\rho]=0$, we have
\begin{eqnarray}
&&
G(\xi_j,\eta_{j\alpha};B)=
\langle
{\rm e}^{C+D}
\Gamma_i(-t_0)^\dag \Gamma_{\tau}^\dag \Gamma_f(\tau)^\dag
{\rm e}^{-2C-2D}\
\Gamma_f(\tau) \Gamma_{\tau} \Gamma_i(-t_0)
\ {\rm e}^{C+D}
\rangle
\ .
\end{eqnarray}
This is our starting point. Note that $C$ and $D$ are Hermitian for real
$\xi_j$ and $\eta_{j\alpha}$ and that $D$ is the function only of $\xi_1-\xi_2$
and $\eta_{1\alpha}-\eta_{2\alpha}$.

\subsection{Some Inequalities}

Here, for the sake of self-containedness, well-known equalities and 
inequalities \cite{BR02}
necessary for the following proof are summarized. For an operator 
$X$, the operator norm
$\Vert X\Vert$ is defined by
\begin{equation}
\Vert X\Vert\equiv \sup_{ |\varphi\rangle\not=0}
\sqrt{\langle \varphi|X^\dag X|\varphi\rangle\over\langle 
\varphi|\varphi\rangle}
\end{equation}
and it satisfies:

\noindent{\bf Equality 1:} For any unitary $U$, $\Vert U^\dag 
XU\Vert=\Vert X\Vert$.

Indeed, we find
\be
\Vert U^\dag XU\Vert^2=\sup_{ |\varphi\rangle\not=0}
{\langle \varphi|U^\dag X^\dag XU|\varphi\rangle\over\langle 
\varphi|\varphi\rangle}
=\sup_{ |\psi\rangle\not=0}
{\langle \psi|X^\dag X|\psi\rangle\over\langle \psi|UU^\dag|\psi\rangle}
=\sup_{ |\psi\rangle\not=0}
{\langle \psi|X^\dag X|\psi\rangle\over\langle \psi|\psi\rangle}
=\Vert X\Vert^2 \ ,
\ee
where we have set $|\psi\rangle=U|\varphi\rangle$.

\noindent{\bf Inequality 1:} $\langle X^\dag Y^\dag YX\rangle \le 
\Vert Y\Vert^2
\langle X^\dag X\rangle$.

Let $\{\varphi_\sigma\}$ be a complete orthonormal basis of eigenvectors
of $\rho$: $\rho|\varphi_\sigma\rangle=\rho_\sigma |\varphi_\sigma\rangle$.
Then, because of $\langle \varphi|X^\dag X|\varphi\rangle
\le \Vert X\Vert^2\langle \varphi|\varphi\rangle$,
\bea
\langle X^\dag Y^\dag YX\rangle &\equiv& \sum_\sigma \rho_\sigma 
\langle  \varphi_\sigma|
X^\dag Y^\dag YX|\varphi_\sigma\rangle \nonumber\\
&\le& \sum_\sigma \rho_\sigma \Vert Y\Vert^2 \langle  \varphi_\sigma|
X^\dag X|\varphi_\sigma\rangle \nonumber\\
&=&\Vert Y\Vert^2 \sum_\sigma \rho_\sigma \langle  \varphi_\sigma|
X^\dag X|\varphi_\sigma\rangle \nonumber\\
&=& \Vert Y\Vert^2
\langle X^\dag X\rangle \ .
\eea

\noindent{\bf Inequality 2:} $\langle X^\dag Y^\dag YX\rangle \le 
\Vert {\rm e}^{-A} X^\dag
{\rm e}^A \Vert^2\langle Y^\dag Y\rangle$ where 
$2A=\sum_j\beta_j(H_j-\sum_\alpha\mu_{j\alpha}N_{j\alpha})$.

Thanks to the cyclicity of the trace, we have the 
Kubo-Martin-Schwinger (KMS) condition
$\langle XY\rangle=\langle {\rm e}^A Y{\rm e}^{-A}{\rm e}^{-A}X{\rm 
e}^A\rangle$ 
for canonical averages $\langle XY\rangle= {1\over 
\Xi}{\rm Tr}({\rm e}^{-2A}XY)$
with $\Xi={\rm Tr}\, {\rm e}^{-2A}$. 
The KMS condition and Inequality 1 imply
\begin{eqnarray}
&&\langle X^\dag Y^\dag YX\rangle = \langle
{\rm e}^A YX{\rm e}^{-A}{\rm e}^{-A}X^\dag Y^\dag {\rm e}^A\rangle
=\langle
{\rm e}^A Y{\rm e}^{-A} {\rm e}^A X{\rm e}^{-A} {\rm e}^{-A} X^\dag 
{\rm e}^A {\rm e}^{-A} Y^\dag {\rm e}^A\rangle
\nonumber\\
&&
\le
\Vert {\rm e}^{-A} X^\dag {\rm e}^A \Vert^2
\langle {\rm e}^A Y{\rm e}^{-A} {\rm e}^{-A} Y^\dag {\rm e}^A\rangle
=\Vert {\rm e}^{-A} X^\dag {\rm e}^A \Vert^2
\langle Y^\dag Y\rangle \; .
\end{eqnarray}

\subsection{Proof}

\noindent{\bf Step 1:} \\
Let $X_1={\rm e}^{-C-D}\Gamma_{\tau} \Gamma_i(-t_0)
\ {\rm e}^{C+D}$. Then Inequality 1 leads to
\begin{eqnarray}
G(\xi_j,\eta_{j\alpha};B) &=&
\langle X_1^\dag \left[
{\rm e}^{-C-D}
\Gamma_f(\tau)
{\rm e}^{C+D}\right]^\dag
\left[
{\rm e}^{-C-D}\,
\Gamma_f(\tau) \, {\rm e}^{C+D}
\right]
X_1
\rangle \nonumber\\
&\le&
\Vert
{\rm e}^{-C-D}
\Gamma_f(\tau) \ {\rm e}^{C+D}
\Vert^2
\langle X_1^\dag
X_1
\rangle
\; .
\end{eqnarray}

Since $\Gamma_f(\tau)^\dag\Gamma_f(\tau)=1$, we have
\begin{eqnarray}
\langle X_1^\dag X_1\rangle &&= \langle X_1^\dag {\rm e}^{C+D}{\rm 
e}^{-C-D}{\rm e}^{-C-D}{\rm e}^{C+D} X_1\rangle \nonumber\\ &&
=\langle X_1^\dag {\rm e}^{C+D} \Gamma_f(\tau)^\dag \Gamma_f(\tau) 
{\rm e}^{-C-D}{\rm e}^{-C-D}
\Gamma_f(\tau)^\dag \Gamma_f(\tau) {\rm e}^{C+D} X_1\rangle
\nonumber\\
&&=\langle X_1^\dag {\rm e}^{C+D} \Gamma_f(\tau)^\dag {\rm 
e}^{-C-D}\left[{\rm e}^{C+D}\Gamma_f(\tau) {\rm e}^{-C-D}\right]
\left[{\rm e}^{-C-D}\Gamma_f(\tau)^\dag {\rm e}^{C+D}\right]{\rm 
e}^{-C-D} \Gamma_f(\tau) {\rm e}^{C+D} X_1\rangle
\nonumber\\
&&=\langle X_1^\dag {\rm e}^{C+D} \Gamma_f(\tau)^\dag {\rm e}^{-C-D} Y_1^\dag
Y_1 \ {\rm e}^{-C-D} \Gamma_f(\tau) {\rm e}^{C+D} X_1\rangle
\nonumber\\
&&\le
\Vert Y_1\Vert^2
\langle X_1^\dag {\rm e}^{C+D} \Gamma_f(\tau)^\dag {\rm e}^{-2C-2D}
\Gamma_f(\tau) {\rm e}^{C+D} X_1\rangle
\nonumber\\
&&\le
\Vert {\rm e}^{-C-D}\Gamma_f(\tau)^\dag {\rm e}^{C+D}\Vert^2
\langle X_1^\dag {\rm e}^{C+D} \Gamma_f(\tau)^\dag {\rm e}^{-2C-2D}
\Gamma_f(\tau) {\rm e}^{C+D} X_1\rangle
\nonumber\\
&&
=\Vert {\rm e}^{-C-D}\Gamma_f(\tau)^\dag {\rm e}^{C+D}\Vert^2
G(\xi_j,\eta_{j\alpha};B)
\; ,
\end{eqnarray}
where $Y_1={\rm e}^{-C-D}\Gamma_f(\tau)^\dag {\rm e}^{C+D}$ and 
Inequality 1 has been used.
\medskip

Since $\Gamma_f(\tau)={\rm e}^{iH_0\tau} \Gamma_f {\rm 
e}^{-iH_0\tau}$ and $[H_0,C]=[H_0,D]=0$,
\bea
\Vert
{\rm e}^{-C-D}
\Gamma_f(\tau) \ {\rm e}^{C+D}
\Vert &=&
\Vert
{\rm e}^{-C-D}{\rm e}^{iH_0\tau}
\Gamma_f {\rm e}^{-iH_0\tau} \ {\rm e}^{C+D}
\Vert \nonumber\\ &=&
\Vert
{\rm e}^{iH_0\tau} {\rm e}^{-C-D}
\Gamma_f {\rm e}^{C+D}
{\rm e}^{-iH_0\tau}
\Vert \nonumber\\ &=&
\Vert
{\rm e}^{-C-D}
\Gamma_f {\rm e}^{C+D}
\Vert \; ,
\eea
where we have used Equality 1 for the norm. Similarly,
$
\Vert {\rm e}^{-C-D}\Gamma_f(\tau)^\dag {\rm e}^{C+D}\Vert
=\Vert {\rm e}^{-C-D}\Gamma_f^\dag {\rm e}^{C+D}\Vert
$.
\medskip

In short, in terms of 
\be
G_1(\xi_j,\eta_{j\alpha};B)\equiv \langle X_1^\dag
\ X_1 \rangle= \langle {\rm e}^{C+D} \Gamma_i(-t_0)^\dag \Gamma_{\tau}^\dag
{\rm e}^{-2C-2D}\Gamma_{\tau} \Gamma_i(-t_0)
\ {\rm e}^{C+D} \rangle \; ,
\ee
we have
\begin{eqnarray}
&&{G_1(\xi_j,\eta_{j\alpha};B)\over
\Vert {\rm e}^{-C-D}\Gamma_f^\dag {\rm e}^{C+D}\Vert^2}
\le G(\xi_j,\eta_{j\alpha};B)
\le
\Vert
{\rm e}^{-C-D}
\Gamma_f \ {\rm e}^{C+D}
\Vert^2
G_1(\xi_j,\eta_{j\alpha};B)\ .
\end{eqnarray}

\noindent{\bf Step 2:} \\
In terms of $X_2={\rm e}^{-C-D}\Gamma_i(-t_0){\rm e}^{C+D}$, one has 
from Inequality 2
\begin{eqnarray}
G_1(\xi_j,\eta_{j\alpha};B) &&= \langle {\rm e}^{C+D} 
\Gamma_i(-t_0)^\dag \Gamma_{\tau}^\dag
{\rm e}^{-2C-2D}\Gamma_{\tau} \Gamma_i(-t_0)
\ {\rm e}^{C+D} \rangle
\nonumber\\
&&= \langle {\rm e}^{C+D} \Gamma_i(-t_0)^\dag {\rm e}^{-C-D} {\rm 
e}^{C+D} \Gamma_{\tau}^\dag
{\rm e}^{-2C-2D}\Gamma_{\tau} {\rm e}^{C+D} {\rm e}^{-C-D} \Gamma_i(-t_0)
\ {\rm e}^{C+D} \rangle
\nonumber\\
&&= \langle X_2^\dag\left[
{\rm e}^{-C-D} \Gamma_{\tau} {\rm e}^{C+D}\right]^\dag
\left[{\rm e}^{-C-D} \Gamma_{\tau} {\rm e}^{C+D} \right]
X_2
\rangle
\nonumber\\
&&\le
\Vert {\rm e}^{-A}
X_2^\dag
{\rm e}^A \Vert^2
\langle \left[{\rm e}^{-C-D} \Gamma_{\tau} {\rm e}^{C+D}\right]^\dag
\left[{\rm e}^{-C-D}\Gamma_{\tau} {\rm e}^{C+D}\right] \rangle
\nonumber\\
&&\le
\Vert {\rm e}^{-A}
{\rm e}^{C+D} \Gamma_i(-t_0)^\dag {\rm e}^{-C-D}
{\rm e}^A \Vert^2
\langle {\rm e}^{C+D} \Gamma_{\tau}^\dag
{\rm e}^{-2C-2D}\Gamma_{\tau} {\rm e}^{C+D} \rangle \; .
\end{eqnarray}
Conversely, in terms of $Y_2={\rm e}^{-C-D} \Gamma_i(-t_0)^\dag {\rm 
e}^{C+D}$,  
Inequality 2 leads to
\begin{eqnarray}
&&\langle {\rm e}^{C+D} \Gamma_{\tau}^\dag
{\rm e}^{-2C-2D}\Gamma_{\tau} {\rm e}^{C+D} \rangle
\nonumber\\
&&=
\langle {\rm e}^{C+D} \Gamma_i(-t_0){\rm e}^{-C-D} {\rm e}^{C+D} 
\Gamma_i(-t_0)^\dag
\Gamma_{\tau}^\dag
{\rm e}^{-2C-2D}\Gamma_{\tau} \Gamma_i(-t_0){\rm e}^{C+D}
{\rm e}^{-C-D}
\Gamma_i(-t_0)^\dag {\rm e}^{C+D} \rangle
\nonumber\\
&&=
\langle Y_2^\dag
{\rm e}^{C+D} \Gamma_i(-t_0)^\dag
\Gamma_{\tau}^\dag
{\rm e}^{-2C-2D}\Gamma_{\tau} \Gamma_i(-t_0){\rm e}^{C+D} Y_2 \rangle
\nonumber\\
&&\le
\Vert {\rm e}^{-A}
Y_2^\dag
{\rm e}^A \Vert^2
\langle {\rm e}^{C+D} \Gamma_i(-t_0)^\dag \Gamma_{\tau}^\dag
{\rm e}^{-2C-2D}\Gamma_{\tau} \Gamma_i(-t_0)
\ {\rm e}^{C+D} \rangle
\nonumber\\
&&\le
\Vert {\rm e}^{-A}
{\rm e}^{C+D} \Gamma_i(-t_0) {\rm e}^{-C-D}
{\rm e}^A \Vert^2
\langle {\rm e}^{C+D} \Gamma_i(-t_0)^\dag \Gamma_{\tau}^\dag
{\rm e}^{-2C-2D}\Gamma_{\tau} \Gamma_i(-t_0)
\ {\rm e}^{C+D} \rangle \; .
\end{eqnarray}
In short, let $G_2(\xi_j,\eta_{j\alpha};B)\equiv\langle {\rm e}^{C+D} 
\Gamma_{\tau}^\dag
{\rm e}^{-2C-2D}\Gamma_{\tau} {\rm e}^{C+D} \rangle$, then
\be
{G_2(\xi_j,\eta_{j\alpha};B)\over \Vert {\rm e}^{-A}
{\rm e}^{C+D} \Gamma_i(-t_0) {\rm e}^{-C-D}
{\rm e}^A \Vert^2}
\le G_1(\xi_j,\eta_{j\alpha};B)\le
\Vert {\rm e}^{-A}
{\rm e}^{C+D} \Gamma_i(-t_0)^\dag {\rm e}^{-C-D}
{\rm e}^A \Vert^2
G_2(\xi_j,\eta_{j\alpha};B) \; .
\ee

\noindent{\bf Step 3:} \\
We set
\begin{eqnarray}
&&2{\bar C}={\xi_1+\xi_2\over 2}{\bar H} +\sum_\alpha 
{\eta_{1\alpha}+\eta_{2\alpha}\over 2}{\bar N}_{\alpha} \; ,
\end{eqnarray}
where ${\bar H}=H_1+H_2+V_0$ and ${\bar 
N}_\alpha=N_{1\alpha}+N_{2\alpha}+n_\alpha$.
Then, in terms of ${\tilde \Gamma}_{\tau}=U_{\tau}{\rm e}^{iH_0\tau}$, we have
\begin{eqnarray}
G_2(\xi_j,\eta_{j\alpha};B) &&=\langle {\rm e}^{C+D}\Gamma_{\tau}^\dag 
{\rm e}^{-2C-2D}
\Gamma_{\tau}{\rm e}^{C+D}\rangle \nonumber\\ &&
=\langle {\rm e}^{C+D}{\rm e}^{iH_0\tau} {\tilde\Gamma}_{\tau}^\dag 
{\rm e}^{-iH_0\tau} {\rm e}^{-2C-2D}
{\rm e}^{iH_0\tau}{\tilde\Gamma}_{\tau}{\rm e}^{-iH_0\tau}{\rm e}^{C+D}\rangle
\nonumber\\
&&=
\langle {\rm e}^{iH_0\tau} {\rm e}^D {\rm e}^C{\tilde 
\Gamma}_{\tau}^\dag {\rm e}^{-{\bar C}}{\rm e}^{-D}{\rm e}^{-iH_0\tau}
\left[{\rm e}^{iH_0\tau}{\rm e}^D{\rm e}^{\bar C}{\rm e}^{-C}{\rm 
e}^{-D}{\rm e}^{-iH_0\tau}\right] \nonumber\\ && \qquad \times
\left[{\rm e}^{iH_0\tau}{\rm e}^{-D}{\rm e}^{-C}{\rm e}^{\bar C}{\rm 
e}^D{\rm e}^{-iH_0\tau}\right]
{\rm e}^{iH_0\tau}{\rm e}^{-D}{\rm e}^{-{\bar C}}{\tilde 
\Gamma}_{\tau}{\rm e}^C{\rm e}^D
{\rm e}^{-iH_0\tau}\rangle
\nonumber\\
&&=
\langle
X_3^\dag
\left[{\rm e}^{iH_0\tau}{\rm e}^{-D}{\rm e}^{-C} {\rm e}^{\bar C} 
{\rm e}^D {\rm e}^{-iH_0\tau}\right]^\dag
\left[{\rm e}^{iH_0\tau}{\rm e}^{-D}{\rm e}^{-C}{\rm e}^{\bar C}{\rm 
e}^D{\rm e}^{-iH_0\tau}\right]
X_3
\rangle
\nonumber\\
&&
\le \Vert {\rm e}^{iH_0\tau}{\rm e}^{-D}{\rm e}^{-C}{\rm e}^{\bar 
C}{\rm e}^D{\rm e}^{-iH_0\tau}\Vert^2
\langle X_3^\dag X_3
\rangle \nonumber\\ &&
=
\Vert
{\rm e}^{-D}{\rm e}^{-C}{\rm e}^{{\bar C}}{\rm e}^D
\Vert^2
\langle X_3^\dag X_3
\rangle  \; ,
\end{eqnarray}
where $X_3={\rm e}^{iH_0\tau}{\rm e}^{-D}{\rm e}^{-{\bar C}}{\tilde 
\Gamma}_{\tau}{\rm e}^C{\rm e}^D
{\rm e}^{-iH_0\tau}$ and $[H_0,C]=[H_0,D]=0$,
Inequality 1 and Equality 1 have been used.
Because of ${\rm e}^{-{\bar C}}{\tilde\Gamma}_{\tau}{\rm e}^C={\rm 
e}^{-iH_0\tau}\Gamma_{\tau}{\rm e}^{-{\bar C}}{\rm e}^C{\rm 
e}^{iH_0\tau}$,
one has
$
X_3={\rm e}^{-D}\Gamma_{\tau}{\rm e}^{-{\bar C}}{\rm e}^C{\rm e}^D
$
and, thus,
\begin{eqnarray}
G_2(\xi_j,\eta_{j\alpha};B)
\le \Vert {\rm e}^{-D}{\rm e}^{-C}{\rm e}^{\bar C}{\rm e}^D\Vert^2
\langle {\rm e}^D{\rm e}^C{\rm e}^{-{\bar C}}\Gamma_{\tau}^\dag {\rm 
e}^{-2D}\Gamma_{\tau}{\rm e}^{-{\bar C}}{\rm e}^C{\rm e}^D
\rangle \; .
\end{eqnarray}

Furthermore, Inequality 2 gives
\begin{eqnarray}
\langle {\rm e}^D{\rm e}^C{\rm e}^{-{\bar C}}\Gamma_{\tau}^\dag 
{\rm e}^{-2D}\Gamma_{\tau}{\rm e}^{-{\bar C}}{\rm e}^C{\rm e}^D\rangle
&&=\langle \left[{\rm e}^{-D}{\rm e}^{-{\bar C}}{\rm e}^C{\rm e}^D\right]^\dag
{\rm e}^D\Gamma_{\tau}^\dag {\rm e}^{-2D}\Gamma_{\tau}{\rm e}^D 
\left[{\rm e}^{-D}{\rm e}^{-{\bar C}}{\rm e}^C{\rm e}^D\right]
\rangle
\nonumber\\
&&\le \Vert {\rm e}^{-A}
\left[{\rm e}^{-D}{\rm e}^{-{\bar C}}{\rm e}^C{\rm e}^D\right]^\dag
{\rm e}^A\Vert^2
\langle {\rm e}^D\Gamma_{\tau}^\dag {\rm e}^{-2D}\Gamma_{\tau}{\rm 
e}^D \rangle \; .
\end{eqnarray}
Thus, $G_3(\xi_j,\eta_{j\alpha};B)\equiv\langle {\rm 
e}^D\Gamma_{\tau}^\dag {\rm e}^{-2D}\Gamma_{\tau}{\rm e}^D \rangle$ 
satisfies
\begin{eqnarray}
G_2(\xi_j,\eta_{j\alpha};B)
\le \Vert {\rm e}^{-D}{\rm e}^{-C}{\rm e}^{\bar C}{\rm e}^D\Vert^2
\Vert {\rm e}^{-A}
\left[{\rm e}^{-D}{\rm e}^{-{\bar C}}{\rm e}^C{\rm e}^D\right]^\dag
{\rm e}^A\Vert^2
G_3(\xi_j,\eta_{j\alpha};B) \; .
\end{eqnarray}

Conversely, we have
\begin{eqnarray}
G_3(\xi_j,\eta_{j\alpha};B) &&=\langle \left[{\rm e}^{-D}{\rm 
e}^{-C}{\rm e}^{\bar C}{\rm e}^D\right]^\dag
{\rm e}^D{\rm e}^C{\rm e}^{-{\bar C}}\Gamma_{\tau}^\dag {\rm 
e}^{-2D}\Gamma_{\tau}{\rm e}^{-{\bar C}}{\rm e}^C{\rm e}^D
\left[{\rm e}^{-D}{\rm e}^{-C}{\rm e}^{\bar C}{\rm e}^D\right]\rangle
\nonumber\\
&&\le \Vert {\rm e}^{-A}
\left[{\rm e}^{-D}{\rm e}^{-C}{\rm e}^{\bar C}{\rm e}^D\right]^\dag
{\rm e}^A\Vert^2
\langle
{\rm e}^D{\rm e}^C{\rm e}^{-{\bar C}}\Gamma_{\tau}^\dag {\rm 
e}^{-2D}\Gamma_{\tau}{\rm e}^{-{\bar C}}{\rm e}^C{\rm e}^D
\rangle
\nonumber\\
&&=\Vert {\rm e}^{-A}
\left[{\rm e}^{-D}{\rm e}^{-C}{\rm e}^{\bar C}{\rm e}^D\right]^\dag
{\rm e}^A\Vert^2
\langle X_3^\dag X_3\rangle \; ,
\end{eqnarray}
where we have used Inequality 2.
Let $Y_3={\rm e}^{iH_0\tau}{\rm e}^{-D}{\rm e}^{-{\bar C}}{\rm 
e}^C{\rm e}^D{\rm e}^{-iH_0\tau}$,
then Inequality 1 and Equality 1 lead to
\begin{eqnarray}
\langle X_3^\dag X_3\rangle
&& =\langle
X_3^\dag
\left[{\rm e}^{iH_0\tau}{\rm e}^{-D}{\rm e}^{-C} {\rm e}^{\bar C} 
{\rm e}^D {\rm e}^{-iH_0\tau}\right]^\dag Y_3^\dag
Y_3\left[{\rm e}^{iH_0\tau}{\rm e}^{-D}{\rm e}^{-C}{\rm e}^{\bar 
C}{\rm e}^D{\rm e}^{-iH_0\tau}\right]
X_3
\rangle
\nonumber\\
&&\le \Vert Y_3\Vert^2
\langle
X_3^\dag
\left[{\rm e}^{iH_0\tau}{\rm e}^{-D}{\rm e}^{-C} {\rm e}^{\bar C} 
{\rm e}^D {\rm e}^{-iH_0\tau}\right]^\dag
\left[{\rm e}^{iH_0\tau}{\rm e}^{-D}{\rm e}^{-C}{\rm e}^{\bar C}{\rm 
e}^D{\rm e}^{-iH_0\tau}\right]
X_3
\rangle
\nonumber\\
&&
=\Vert {\rm e}^{iH_0\tau}{\rm e}^{-D}{\rm e}^{-{\bar C}}{\rm e}^C{\rm 
e}^D{\rm e}^{-iH_0\tau}\Vert^2
G_2(\xi_j,\eta_{j\alpha};B) \nonumber\\ &&
=\Vert {\rm e}^{-D}{\rm e}^{-{\bar C}}{\rm e}^C{\rm e}^D\Vert^2
G_2(\xi_j,\eta_{j\alpha};B) \; .
\end{eqnarray}
Thus,
\begin{eqnarray}
&&{G_3(\xi_j,\eta_{j\alpha};B)\over
\Vert {\rm e}^{-A}
\left[{\rm e}^{-D}{\rm e}^{-C}{\rm e}^{\bar C}{\rm e}^D\right]^\dag
{\rm e}^A\Vert^2
\Vert {\rm e}^{-D}{\rm e}^{-{\bar C}}{\rm e}^C{\rm e}^D\Vert^2}
\le
G_2(\xi_j,\eta_{j\alpha};B) \; .
\end{eqnarray}

\noindent{\bf Step 4:} \\
From Steps 1 to 3, in terms of
\begin{eqnarray}
L&=&{1\over
\Vert {\rm e}^{-C-D}\Gamma_f^\dag {\rm e}^{C+D}\Vert^2
\Vert {\rm e}^{-A}
{\rm e}^{C+D} \Gamma_i(-t_0) {\rm e}^{-C-D}
{\rm e}^A \Vert^2
\Vert {\rm e}^{-A}
\left[{\rm e}^{-D}{\rm e}^{-C}{\rm e}^{\bar C}{\rm e}^D\right]^\dag
{\rm e}^A\Vert^2
\Vert {\rm e}^{-D}{\rm e}^{-{\bar C}}{\rm e}^C{\rm e}^D\Vert^2
} \; , \nonumber\\ &&
\label{constL}
\\
K&=&\Vert {\rm e}^{-D}{\rm e}^{-C}{\rm e}^{\bar C}{\rm e}^D\Vert^2
\Vert {\rm e}^{-A}
\left[{\rm e}^{-D}{\rm e}^{-{\bar C}}{\rm e}^C{\rm e}^D\right]^\dag
{\rm e}^A\Vert^2
\Vert {\rm e}^{-A}
{\rm e}^{C+D} \Gamma_i(-t_0)^\dag {\rm e}^{-C-D}
{\rm e}^A \Vert^2
\Vert
{\rm e}^{-C-D}
\Gamma_f \ {\rm e}^{C+D}
\Vert^2 \; , \nonumber\\ &&
\label{constK}
\end{eqnarray}
we have
\begin{eqnarray}
&&
L\
G_3(\xi_j,\eta_{j\alpha};B)
\le G(\xi_j,\eta_{j\alpha};B)\le
K\ G_3(\xi_j,\eta_{j\alpha};B)
\ .
\end{eqnarray}
Note that the constants $L$ and $K$ are independent of $\tau$ and that
$G_3(\xi_j,\eta_{j\alpha};B)$ is a function only of $\xi_1-\xi_2$
and $\eta_{1\alpha}-\eta_{2\alpha}$ since the operator $D$ depends
only on them.

\medskip

\noindent{\bf Step 5:}\\
If
\be
Q(\xi_j,\eta_{j\alpha};B)\equiv -
\lim_{{\cal T}\to\infty}{1\over {\cal T}}\ln G(\xi_j,\eta_{j\alpha};B)
\ee
exists, one has
\begin{eqnarray}
Q(\xi_j,\eta_{j\alpha};B) && =
-\lim_{{\cal T}\to\infty}{1\over{\cal T}}\ln G(\xi_j,\eta_{j\alpha};B)
+\lim_{{\cal T}\to\infty}{1\over{\cal T}}\ln L\nonumber\\ && \le
-\lim_{{\cal T}\to\infty}{1\over{\cal T}}\ln G_3(\xi_j,\eta_{j\alpha};B)
\nonumber\\
&&\le
-\lim_{{\cal T}\to\infty}{1\over{\cal T}}\ln G(\xi_j,\eta_{j\alpha};B)
+\lim_{{\cal T}\to\infty}{1\over{\cal T}}\ln K
=Q(\xi_j,\eta_{j\alpha};B) \; .
\end{eqnarray}
In short, we have shown:
\begin{eqnarray}
Q(\xi_j,\eta_{j\alpha};B)= -\lim_{{\cal T}\to\infty}{1\over{\cal 
T}}\ln G_3(\xi_j,\eta_{j\alpha};B)
=-\lim_{{\cal T}\to\infty}{1\over{\cal T}}\ln
\langle {\rm e}^D\Gamma_{\tau}^\dag {\rm e}^{-D} {\rm e}^{-D} 
\Gamma_{\tau} {\rm e}^D \rangle  \; .
\label{explicit}
\end{eqnarray}
The left-most term only contains $D$, which is a function only of 
$\xi_1-\xi_2$ and
$\eta_{1\alpha}-\eta_{2\alpha}$, or
\be
Q(\xi_j,\eta_{j\alpha};B)= {\widetilde 
Q}(\xi_1-\xi_2,\eta_{1\alpha}-\eta_{2\alpha};B) \; .
\label{Q=Qtilde}
\ee
{\it Q.E.D.}

We would like to remark that, even when the system has very long but 
finite recurrent times, 
the quantities $Q$ and $\widetilde Q$ can be 
introduced and the Proposition is expected to hold 
with errors of 
order of $1/{\cal T}$.
Firstly, in such a case, the ratio $-\ln G/{\cal T}$ would converge 
to a definite value $Q$ provided ${\cal T}$ is sufficiently longer 
than the relaxation time but shorter than the recurrent time. 
The 
same would be valid for ${\widetilde Q}$.
Secondly, even in such a case, the quantities appearing in 
(\ref{constL}) 
and (\ref{constK}) are bounded by constants 
independent 
of the reservoir volumes $\Omega$ if the interaction 
$V(t)$ and the subsystem particle
numbers $n_\alpha$ have finite norms.
Indeed, $\Gamma_i(-t_0)={\rm e}^{-iH_0t_0}{\rm T} \exp\left[-i\int_0^{t_0}
{\rm e}^{iH_0s}V(s){\rm e}^{-iH_0s}ds\right]{\rm e}^{iH_0t_0}$ 
where 
`T~$\exp$' is the time-ordered exponential,
and its norm is bounded by $\exp\left[t_0\sup_t\Vert 
V(t)\Vert\right]$~\cite{BR02}.
Then, the product ${\rm e}^{C+D}\Gamma_i(-t_0){\rm e}^{-C-D}$ is 
$\Omega$-independent and has a finite norm because ${\rm e}^{-C-D}$ 
is O$({\rm e}^{\mp\Omega})$ 
if ${\rm e}^{C+D}$ is O$({\rm 
e}^{\pm\Omega})$.
By similar arguments, all the norms in Eqs. (\ref{constL}) and 
(\ref{constK}) 
are found to be $\Omega$-independent and, thus,
the difference $Q-{\widetilde Q}$ is of the order of $1/{\cal T}$ 
instead of $\Omega/{\cal T}$.
Accordingly, the equality 
(\ref{Q=Qtilde}) is obtained in the limit ${\cal T}\to\infty$.

\subsection{Generalization}

The previous results can be generalized to the case of $r>2$ reservoirs.
In this case, the proposition (\ref{dfn-tilde-Q}) is that the generating function
is a function 
\be
Q(\xi_j,\eta_{j\alpha};B)=\Q(\tilde\xi_j,\tilde\eta_{j\alpha};B) \; ,
\ee
depending only on the independent parameters:
\bea
{\tilde \xi}_j &\equiv& \xi_j-\frac{1}{r}\sum_{k=1}^r \xi_k  \; ,\\
{\tilde \eta}_{j\alpha} &\equiv& \eta_{j\alpha}-\frac{1}{r}\sum_{k=1}^r \eta_{k\alpha} \; ,
\eea
with $j=1,2,...,r-1$. The proof is similar as in the case $r=2$ with the operators:
\bea
&&2C=\frac{1}{r}\sum_{k=1}^r \left( \xi_k H_0 +\sum_{\alpha} \eta_{k\alpha} N_{0\alpha}\right) \; ,
\\
&&2D=\sum_{k=1}^r \left( \tilde\xi_k H_k + \sum_{\alpha} \tilde\eta_{k\alpha} N_{k\alpha}\right)\; ,
\eea
replacing Eqs. (\ref{C})-(\ref{D}), where $H_0=\sum_{k=1}^r H_k$ and $N_{0\alpha}=\sum_{k=1}^r N_{k\alpha}$.

In the general case, the fluctuation theorem should read
\be
{\tilde Q}({\tilde\xi}_j,{\tilde\eta}_{j\alpha};B)={\tilde Q}({\tilde A}_{j0}-{\tilde\xi}_j,{\tilde A}_{j\alpha}-{\tilde\eta}_{j\alpha};-B) \; ,
\ee
in terms of the independent affinities
\bea
{\tilde A}_{j0} &\equiv& \beta_j  - \frac{1}{r}\sum_{k=1}^r \beta_k \; ,\label{A_j0}\\
{\tilde A}_{j\alpha} &\equiv& -\beta_j\mu_{j\alpha}+ \frac{1}{r}\sum_{k=1}^r\beta_k\mu_{k\alpha} \; , \qquad\mbox{for} \quad \alpha=1,2,...,c,
\label{A_jalpha}
\eea
with $j=1,2,...,r-1$.

\section{Symmetry relations for the response coefficients}
\label{NLR}

\subsection{Fluctuation theorem and response coefficients}

If we gather the independent parameters and affinities in the case of $r=2$ reservoirs as
\bea
\pmb{\lambda}&=& \{ \xi_1-\xi_2, \eta_{1\alpha}-\eta_{2\alpha} \} \; , \\
\pmb{A}&=& \{ A_0, A_{\alpha}\} \; ,
\eea
or in the general case of $r>2$ reservoirs as
\bea
\pmb{\lambda}&=& \{ {\tilde\xi}_j, {\tilde\eta}_{j\alpha} \} \; ,\\
\pmb{A}&=& \{ {\tilde A}_{j0}, {\tilde A}_{j\alpha}\} \; ,
\eea
with $\alpha=1,2,...,c$ and $j=1,2,...,r-1$,
the fluctuation theorem (\ref{FT}) reads
\be
\Q(\pmb{\lambda},\pmb{A};B) = \Q(\pmb{A}-\pmb{\lambda},\pmb{A};-B) \; ,
\ee
where we have explicitly written the dependence of the generating function on the affinities
defining the nonequilibrium steady state.

The idea is to differentiate successively the fluctuation theorem with respect to
both $\pmb{\lambda}$ and $\pmb{A}$ to obtain symmetry relations 
for the linear and nonlinear response coefficients as well as further coefficients
characterizing the statistics of the current fluctuations \cite{AG04}.

On the one hand, the mean currents can be obtained from the generating function and,
on the other hand, expanded in powers of the affinities:
\begin{equation}
J_{\alpha}(B) \equiv  \frac{\partial \Q}{\partial
\lambda_{\alpha}}(\pmb{0},\pmb{A};B) =
\sum_{\beta} L_{\alpha,\beta}(B) A_{\beta} + \frac{1}{2}
\sum_{\beta,\gamma} M_{\alpha,\beta\gamma}(B) A_{\beta}
A_{\gamma}
+ \frac{1}{6} \sum_{\beta,\gamma,\delta} N_{\alpha,\beta\gamma\delta}(B)
A_{\beta} A_{\gamma} A_{\delta} + \cdots \; ,
\label{dfn-currents}
\end{equation}
which defines the response coefficients:
\bea
L_{\alpha,\beta}(B) &\equiv& \frac{\partial^2 \Q}{\partial \lambda_{\alpha}\partial A_{\beta}}(\pmb{0},\pmb{0};B) \; ,\\
M_{\alpha,\beta\gamma}(B) &\equiv& \frac{\partial^3 \Q}{\partial \lambda_{\alpha}\partial A_{\beta}\partial A_{\gamma}}(\pmb{0},\pmb{0};B) \; ,\label{M}\\
N_{\alpha,\beta\gamma\delta}(B) &\equiv& \frac{\partial^4 \Q}{\partial \lambda_{\alpha}\partial A_{\beta}\partial A_{\gamma}\partial A_{\delta}}(\pmb{0},\pmb{0};B) \; , \label{N}\\
&& \nonumber \\
&&\cdots \nonumber
\eea
around the state of thermodynamic equilibrium.

We notice that, if we set $\pmb{\lambda}=\pmb{0}$ in the fluctuation theorem (\ref{FT}), 
we obtain the identities
\bea
\Q(\pmb{0},\pmb{A};B) &=& 0 \; , \label{norm}\\
\Q(\pmb{A},\pmb{A};-B) &=& 0 \; .\label{global}
\eea
The former is a condition of normalization and the latter a condition of global detailed balancing which is a consequence of the fluctuation theorem (\ref{FT}) but may be assumed for itself as a weaker property than the fluctuation theorem  \cite{FB08}. On the other hand, the generating function of the cumulants
of the fluctuating currents at equilibrium satisfies
\be
\Q(\pmb{\lambda},\pmb{0};B) = \Q(-\pmb{\lambda},\pmb{0};-B) \; ,
\label{FT-eq}
\ee
obtained from the fluctuation theorem (\ref{FT}) at the equilibrium $\pmb{A}=\pmb{0}$.

We start by differentiating the fluctuation theorem with respect to the generating parameters $\{\lambda_{\alpha}\}$ and also
the affinities $\{A_{\alpha}\}$ to get
\bea
\frac{\partial \Q}{\partial \lambda_{\alpha}}(\pmb{\lambda},\pmb{A};B) &=& - \frac{\partial \Q}{\partial \lambda_{\alpha}}(\pmb{A}-\pmb{\lambda},\pmb{A};-B) \; ,\label{1}\\
\frac{\partial \Q}{\partial A_{\alpha}}(\pmb{\lambda},\pmb{A};B) &=& \frac{\partial \Q}{\partial \lambda_{\alpha}}(\pmb{A}-\pmb{\lambda},\pmb{A};-B) + \frac{\partial \Q}{\partial A_{\alpha}}(\pmb{A}-\pmb{\lambda},\pmb{A};-B) \; .\label{2}
\eea
We set $\pmb{\lambda}=\pmb{0}$ in Eq. (\ref{2}), use the conditions (\ref{norm})-(\ref{global}), and set $\pmb{A}=\pmb{0}$ to get
\bea
\frac{\partial \Q}{\partial \lambda_{\alpha}}(\pmb{0},\pmb{0};B) &=& 0 \; ,\\
\frac{\partial \Q}{\partial A_{\alpha}}(\pmb{0},\pmb{0};B) &=& 0 \; ,
\eea
which shows in particular that the mean currents vanish at equilibrium.

\subsection{Onsager-Casimir reciprocity relations and Green-Kubo formulas}

Now, we differentiate Eq. (\ref{1}) with respect to $\lambda_{\beta}$ to obtain
\be
\frac{\partial^2 \Q}{\partial \lambda_{\alpha}\partial \lambda_{\beta}}(\pmb{\lambda},\pmb{A};B) = \frac{\partial^2 \Q}{\partial \lambda_{\alpha}\partial \lambda_{\beta}}(\pmb{A}-\pmb{\lambda},\pmb{A};-B) \; . \label{11}
\ee
Setting $\pmb{\lambda}=\pmb{A}=\pmb{0}$, we find the identity
\be
\frac{\partial^2 \Q}{\partial \lambda_{\alpha}\partial \lambda_{\beta}}(\pmb{0},\pmb{0};B) = \frac{\partial^2 \Q}{\partial \lambda_{\alpha}\partial \lambda_{\beta}}(\pmb{0},\pmb{0};-B) \; ,\label{11-0}
\ee
for the second-order cumulant of the current fluctuations at equilibrium.
 
If we differentiate Eq. (\ref{1}) with respect to $A_{\beta}$, we get
\be
\frac{\partial^2 \Q}{\partial \lambda_{\alpha}\partial A_{\beta}}(\pmb{\lambda},\pmb{A};B) = -\frac{\partial^2 \Q}{\partial \lambda_{\alpha}\partial \lambda_{\beta}}(\pmb{A}-\pmb{\lambda},\pmb{A};-B)-\frac{\partial^2 \Q}{\partial \lambda_{\alpha}\partial A_{\beta}}(\pmb{A}-\pmb{\lambda},\pmb{A};-B) \; ,\label{12}
\ee
which reduces to
\be
L_{\alpha,\beta}(B)  = - \frac{\partial^2 \Q}{\partial \lambda_{\alpha}\partial \lambda_{\beta}}(\pmb{0},\pmb{0};B) 
-L_{\alpha,\beta}(-B) \; ,\label{12-0}
\ee
for $\pmb{\lambda}=\pmb{A}=\pmb{0}$.  We recover the formulas of Green-Kubo type in the case $\alpha=\beta$:
\be
L_{\alpha,\alpha}(B) = - \frac{1}{2}\frac{\partial^2 \Q}{\partial \lambda_{\alpha}^2}(\pmb{0},\pmb{0};B) \; . \label{Kubo}
\ee

The differentiation of Eq. (\ref{2}) with respect to $A_{\beta}$ leads to
\bea
\frac{\partial^2 \Q}{\partial A_{\alpha}\partial A_{\beta}}(\pmb{\lambda},\pmb{A};B) &=& \frac{\partial^2 \Q}{\partial \lambda_{\alpha}\partial \lambda_{\beta}}(\pmb{A}-\pmb{\lambda},\pmb{A};-B)+
\frac{\partial^2 \Q}{\partial \lambda_{\alpha}\partial A_{\beta}}(\pmb{A}-\pmb{\lambda},\pmb{A};-B)
\nonumber\\ &+&
\frac{\partial^2 \Q}{\partial A_{\alpha}\partial \lambda_{\beta}}(\pmb{A}-\pmb{\lambda},\pmb{A};-B)+
\frac{\partial^2 \Q}{\partial A_{\alpha}\partial A_{\beta}}(\pmb{A}-\pmb{\lambda},\pmb{A};-B) \; .\label{22}
\eea
Using Eqs. (\ref{norm}) and (\ref{global}) in the limit $\pmb{\lambda}=\pmb{A}=\pmb{0}$, we have
\be
0 = \frac{\partial^2 \Q}{\partial \lambda_{\alpha}\partial \lambda_{\beta}}(\pmb{0},\pmb{0};-B)+
L_{\alpha,\beta}(-B)+L_{\beta,\alpha}(-B) \; .\label{22-0}
\ee
Combining with Eq. (\ref{12-0}), we finally find the Onsager-Casimir reciprocity relations:
\be
L_{\alpha,\beta}(B)=L_{\beta,\alpha}(-B) \; .\label{OC}
\ee
We notice that Eq. (\ref{12-0}) leading to the Onsager-Casimir relation requires the link established 
by the fluctuation theorem (\ref{FT}) between the variables $\pmb{\lambda}$ and $\pmb{A}$
and does not result from Eqs. (\ref{norm}), (\ref{global}), and (\ref{FT-eq}) alone.

\subsection{Symmetry relations at 2nd order}

We proceed in a similar way to obtain relations at second order.  The differentiation of Eq. (\ref{11}) with respect to $\lambda_{\gamma}$ gives
\be
\frac{\partial^3 \Q}{\partial \lambda_{\alpha}\partial \lambda_{\beta}\partial \lambda_{\gamma}}(\pmb{\lambda},\pmb{A};B) = -\frac{\partial^3 \Q}{\partial \lambda_{\alpha}\partial \lambda_{\beta}\partial \lambda_{\gamma}}(\pmb{A}-\pmb{\lambda},\pmb{A};-B) \; ,\label{111}
\ee
which reduces to
\be
\frac{\partial^3 \Q}{\partial \lambda_{\alpha}\partial \lambda_{\beta}\partial \lambda_{\gamma}}(\pmb{0},\pmb{0};B) = -\frac{\partial^3 \Q}{\partial \lambda_{\alpha}\partial \lambda_{\beta}\partial \lambda_{\gamma}}(\pmb{0},\pmb{0};-B) \; ,\label{111-0}
\ee
for $\pmb{\lambda}=\pmb{A}=\pmb{0}$.

On the other hand, the differentiation of Eq. (\ref{11}) with respect to $A_{\gamma}$ gives
\be
\frac{\partial^3 \Q}{\partial \lambda_{\alpha}\partial \lambda_{\beta}\partial A_{\gamma}}(\pmb{\lambda},\pmb{A};B) = \frac{\partial^3 \Q}{\partial \lambda_{\alpha}\partial \lambda_{\beta}\partial \lambda_{\gamma}}(\pmb{A}-\pmb{\lambda},\pmb{A};-B) + \frac{\partial^3 \Q}{\partial \lambda_{\alpha}\partial \lambda_{\beta}\partial A_{\gamma}}(\pmb{A}-\pmb{\lambda},\pmb{A};-B) \; .\label{112}
\ee
Here, we introduce the coefficients
\be
R_{\alpha\beta,\gamma}(B) \equiv -\frac{\partial^3 \Q}{\partial \lambda_{\alpha}\partial \lambda_{\beta}\partial A_{\gamma}}(\pmb{0},\pmb{0};B) \; ,
\ee
which characterizes the nonequilibrium response of the second cumulants of the current fluctuations.
Setting $\pmb{\lambda}=\pmb{A}=\pmb{0}$ in Eq. (\ref{112}) leads to the relation
\be
R_{\alpha\beta,\gamma}(B)-R_{\alpha\beta,\gamma}(-B)=\frac{\partial^3 \Q}{\partial \lambda_{\alpha}\partial \lambda_{\beta}\partial \lambda_{\gamma}}(\pmb{0},\pmb{0};B) \; .
\label{112-0}
\ee

Now, if we differentiate Eq. (\ref{12}) with respect to $A_{\gamma}$, we get
\bea
\frac{\partial^3 \Q}{\partial \lambda_{\alpha}\partial A_{\beta}\partial A_{\gamma}}(\pmb{\lambda},\pmb{A};B) &=& -\frac{\partial^3 \Q}{\partial \lambda_{\alpha}\partial \lambda_{\beta}\partial\lambda_{\gamma}}(\pmb{A}-\pmb{\lambda},\pmb{A};-B)-\frac{\partial^3 \Q}{\partial \lambda_{\alpha}\partial \lambda_{\beta}\partial A_{\gamma}}(\pmb{A}-\pmb{\lambda},\pmb{A};-B)
\nonumber \\ && -\frac{\partial^3 \Q}{\partial \lambda_{\alpha}\partial A_{\beta}\partial\lambda_{\gamma}}(\pmb{A}-\pmb{\lambda},\pmb{A};-B)-\frac{\partial^3 \Q}{\partial \lambda_{\alpha}\partial A_{\beta}\partial A_{\gamma}}(\pmb{A}-\pmb{\lambda},\pmb{A};-B) \; ,
\nonumber\\ && \label{122}
\eea
whereupon we find for $\pmb{\lambda}=\pmb{A}=\pmb{0}$ that
\be
M_{\alpha,\beta\gamma}(B)+M_{\alpha,\beta\gamma}(-B)=R_{\alpha\beta,\gamma}(-B)+R_{\alpha\gamma,\beta}(-B)+
\frac{\partial^3 \Q}{\partial \lambda_{\alpha}\partial \lambda_{\beta}\partial \lambda_{\gamma}}(\pmb{0},\pmb{0};B)\; ,
\label{122-0}
\ee
involving the second-order response coefficient (\ref{M}).

We end with the differentiation of Eq. (\ref{22}) with respect to $A_{\gamma}$ to obtain
\bea
\frac{\partial^3 \Q}{\partial A_{\alpha}\partial A_{\beta}\partial A_{\gamma}}(\pmb{\lambda},\pmb{A};B) &=& \frac{\partial^3 \Q}{\partial \lambda_{\alpha}\partial \lambda_{\beta}\partial\lambda_{\gamma}}(\pmb{A}-\pmb{\lambda},\pmb{A};-B) +\frac{\partial^3 \Q}{\partial \lambda_{\alpha}\partial \lambda_{\beta}\partial A_{\gamma}}(\pmb{A}-\pmb{\lambda},\pmb{A};-B)
\nonumber \\ &+& \frac{\partial^3 \Q}{\partial \lambda_{\alpha}\partial A_{\beta}\partial\lambda_{\gamma}}(\pmb{A}-\pmb{\lambda},\pmb{A};-B) +\frac{\partial^3 \Q}{\partial \lambda_{\alpha}\partial A_{\beta}\partial A_{\gamma}}(\pmb{A}-\pmb{\lambda},\pmb{A};-B) 
\nonumber \\ &+& \frac{\partial^3 \Q}{\partial A_{\alpha}\partial \lambda_{\beta}\partial\lambda_{\gamma}}(\pmb{A}-\pmb{\lambda},\pmb{A};-B) +\frac{\partial^3 \Q}{\partial A_{\alpha}\partial \lambda_{\beta}\partial A_{\gamma}}(\pmb{A}-\pmb{\lambda},\pmb{A};-B) 
\nonumber \\ &+& \frac{\partial^3 \Q}{\partial A_{\alpha}\partial A_{\beta}\partial\lambda_{\gamma}}(\pmb{A}-\pmb{\lambda},\pmb{A};-B) +\frac{\partial^3 \Q}{\partial A_{\alpha}\partial A_{\beta}\partial A_{\gamma}}(\pmb{A}-\pmb{\lambda},\pmb{A};-B) \; .
\nonumber\\ &&
\label{222}
\eea
Taking $\pmb{\lambda}=\pmb{A}=\pmb{0}$ therein, we deduce
\be
M_{\alpha,\beta\gamma}(B)+M_{\beta,\gamma\alpha}(B)+M_{\gamma,\alpha\beta}(B)=
R_{\alpha\beta,\gamma}(B)+R_{\beta\gamma,\alpha}(B)+R_{\gamma\alpha,\beta}(B)
-\frac{\partial^3 \Q}{\partial \lambda_{\alpha}\partial \lambda_{\beta}\partial \lambda_{\gamma}}(\pmb{0},\pmb{0};B)\; .
\label{222-0}
\ee
We notice that this relation is the consequence of the weaker condition of global detailed balancing (\ref{global}) alone and could hold even though the fluctuation theorem (\ref{FT}) does not as recently shown in Ref. \cite{FB08} where the versions of Eq. (\ref{222-0}), which are (anti)symmetrized with respect to the magnetic field, appears with the notations $M_{\alpha,\beta\gamma}=(k_{\rm B}T)^2 G_{\alpha,\beta\gamma}^{(2)}$, $R_{\alpha\beta,\gamma}=k_{\rm B}T S_{\alpha\beta,\gamma}^{(1)}$, and $\frac{\partial^3 \Q}{\partial \lambda_{\alpha}\partial \lambda_{\beta}\partial \lambda_{\gamma}}(\pmb{0},\pmb{0};B)=C_{\alpha\beta\gamma}^{(0)}$. We point out that 
Eqs. (\ref{112-0}) and (\ref{122-0}) are on the other hand consequences of microreversibility and the stronger fluctuation theorem, as it is the case for the Onsager-Casimir reciprocity relations.

Adding Eq. (\ref{222-0}) to the same equation with $-B$ instead of $B$ and using Eqs. (\ref{122-0}), we moreover infer that
\be
\frac{\partial^3 \Q}{\partial \lambda_{\alpha}\partial \lambda_{\beta}\partial \lambda_{\gamma}}(\pmb{0},\pmb{0};B) = 0 \; ,
\label{111-0-0}
\ee
whereupon we finally find the symmetry relations:
\bea
R_{\alpha\beta,\gamma}(B) &=& R_{\alpha\beta,\gamma}(-B) \; ,\label{R-R}\\
M_{\alpha,\beta\gamma}(B)+M_{\alpha,\beta\gamma}(-B) &=& R_{\alpha\beta,\gamma}(B)+R_{\alpha\gamma,\beta}(B) \; ,\label{M-R}\\
M_{\alpha,\beta\gamma}(B)+M_{\beta,\gamma\alpha}(B)+M_{\gamma,\alpha\beta}(B) &=&
R_{\alpha\beta,\gamma}(B)+R_{\beta\gamma,\alpha}(B)+R_{\gamma\alpha,\beta}(B)\; .
\label{MMM}
\eea
If the magnetic field vanishes $B=0$, Eqs. (\ref{M-R}) reduces to the symmetry relations
\be
M_{\alpha,\beta\gamma}(0)= \frac{1}{2} \left[ R_{\alpha\beta,\gamma}(0)+R_{\alpha\gamma,\beta}(0)\right]\; ,
\ee
which were previously deduced as consequences of the fluctuation theorem \cite{AG06JSM,AG07JSM}. 

\subsection{Symmetry relations at 3rd order}

Beside the third-order response coefficient (\ref{N}), we introduce the coefficients
with
\begin{equation}
S_{\alpha\beta\gamma,\delta}(B) \equiv
\frac{\partial^4 Q}{\partial \lambda_{\alpha}\partial
\lambda_{\beta}\partial\lambda_{\gamma}\partial A_{\delta}}(\pmb{0},\pmb{0};B)
\label{S}
\end{equation}
and
\begin{equation}
T_{\alpha\beta,\gamma\delta}(B) \equiv
- \frac{\partial^4 Q}{\partial \lambda_{\alpha}\partial
\lambda_{\beta}\partial A_{\gamma}\partial
A_{\delta}}(\pmb{0},\pmb{0};B)\; ,
\label{T}
\end{equation}
characterizing the nonequilibrium responses of respectively the second and third cumulants of the current fluctuations.

We continue the deduction by further differentiating the relations of the previous subsection with respect to $\lambda_{\delta}$ or $A_{\delta}$ at $\pmb{\lambda}=\pmb{A}=\pmb{0}$.
We find successively from Eq. (\ref{111}):
\be
\frac{\partial^4 \Q}{\partial \lambda_{\alpha}\partial \lambda_{\beta}\partial \lambda_{\gamma}\partial \lambda_{\delta}}(\pmb{0},\pmb{0};B) = \frac{\partial^4 \Q}{\partial \lambda_{\alpha}\partial \lambda_{\beta}\partial \lambda_{\gamma}\partial \lambda_{\delta}}(\pmb{0},\pmb{0};-B) \label{1111-0}
\ee
and
\be
S_{\alpha\beta\gamma,\delta}(B)+S_{\alpha\beta\gamma,\delta}(-B)=-\frac{\partial^4 \Q}{\partial \lambda_{\alpha}\partial \lambda_{\beta}\partial \lambda_{\gamma}\partial \lambda_{\delta}}(\pmb{0},\pmb{0};B)\; ,
\label{1112-0}
\ee
from Eq. (\ref{112}):
\be
T_{\alpha\beta,\gamma\delta}(B) - T_{\alpha\beta,\gamma\delta}(-B) 
= S_{\alpha\beta\gamma,\delta}(B) - S_{\alpha\beta\delta,\gamma}(-B)\; ,
\label{1122-0}
\ee
from Eq. (\ref{122}):
\bea
N_{\alpha,\beta\gamma\delta}(B) + N_{\alpha,\beta\gamma\delta}(-B) &=& 
T_{\alpha\beta,\gamma\delta}(-B) + T_{\alpha\gamma,\beta\delta}(-B) + T_{\alpha\delta,\beta\gamma}(-B) 
\nonumber \\ &-&
S_{\alpha\beta\gamma,\delta}(-B) -S_{\alpha\beta\delta,\gamma}(-B) -S_{\alpha\gamma\delta,\beta}(-B) 
\nonumber\\ &-& \frac{\partial^4 \Q}{\partial \lambda_{\alpha}\partial \lambda_{\beta}\partial \lambda_{\gamma}\partial \lambda_{\delta}}(\pmb{0},\pmb{0};B) \; ,
\label{1222-0}
\eea
and from Eq. (\ref{222}):
\bea
&& N_{\alpha,\beta\gamma\delta}(B) + N_{\beta,\gamma\delta\alpha}(B)+N_{\gamma,\delta\alpha\beta}(B) + N_{\delta,\alpha\beta\gamma}(B) \nonumber \\ &=&
T_{\alpha\beta,\gamma\delta}(B) + T_{\alpha\gamma,\beta\delta}(B) + T_{\alpha\delta,\beta\gamma}(B) 
+T_{\beta\gamma,\alpha\delta}(B) + T_{\beta\delta,\alpha\gamma}(B) + T_{\gamma\delta,\alpha\beta}(B)
\nonumber \\ &-&
S_{\alpha\beta\gamma,\delta}(B) -S_{\beta\gamma\delta,\alpha}(B)-S_{\gamma\delta\alpha,\beta}(B)-S_{\delta\alpha\beta,\gamma}(B)
\nonumber\\ &-&
\frac{\partial^4 \Q}{\partial \lambda_{\alpha}\partial \lambda_{\beta}\partial \lambda_{\gamma}\partial \lambda_{\delta}}(\pmb{0},\pmb{0};B) \; .
\label{2222-0}
\eea
This last relation is the consequence of the weaker condition of global detailed balancing in the same way as Eq. (\ref{222-0}).

In the case of a vanishing magnetic field $B=0$, Eqs. (\ref{1112-0}) and (\ref{1222-0}) reduce respectively to
\be
S_{\alpha\beta\gamma,\delta}(0)= -\frac{1}{2} \frac{\partial^4 \Q}{\partial \lambda_{\alpha}\partial \lambda_{\beta}\partial \lambda_{\gamma}\partial \lambda_{\delta}}(\pmb{0},\pmb{0};0) 
\label{1112-0-0}
\ee
and
\be
N_{\alpha,\beta\gamma\delta}(0) = \frac{1}{2}\left[
T_{\alpha\beta,\gamma\delta}(0) + T_{\alpha\gamma,\beta\delta}(0) + T_{\alpha\delta,\beta\gamma}(0) 
-S_{\alpha\beta\gamma,\delta}(0) \right]  \; ,
\label{1222-0-0}
\ee
which has been obtained elsewhere as a consequence of the fluctuation theorem \cite{AG07JSM}.

\section{Conclusions}
\label{Conclusions}

In this paper, a fluctuation theorem for the currents has been proved 
for open quantum systems reaching a nonequilibrium steady state in 
the long-time limit \cite{our_story}. In the considered protocol, 
the heat and particle currents are defined in 
terms of the exchanges of energy and particles between reservoirs,
as measured at the initial and final times when the reservoirs are decoupled.

We start from a general symmetry relation for the generating function 
of the exchanges of energy and particles, which is the consequence of 
the microreversibility guaranteed by the measurement protocol. 
This symmetry relation is expressed in terms of 
the temperatures and chemical potentials of the reservoirs.  However, 
the fluctuation theorem for the currents requires a symmetry with 
respect to the thermodynamic forces or affinities which are given in 
terms of the differences of temperatures and chemical potentials.

We show that this symmetry indeed holds by proving that, in the 
long-time limit, the generating function only depends on the {\it 
differences} between the parameters corresponding to the different 
reservoirs.
Accordingly, the generating  function of the currents has the 
symmetry of the fluctuation theorem with respect to the affinities.
A rigorous proof is thus established for 
such steady-state fluctuation theorems 
as considered in the review \cite{EHM08}.

As a consequence, the Onsager-Casimir reciprocity relations can be 
obtained for the linear response coefficients from the fluctuation 
theorem. Furthermore, generalizations of the reciprocity relations to 
the nonlinear response coefficients can also be deduced.

\vskip 0.3 cm

{\bf Acknowledgments.} D.~Andrieux thanks the F.R.S.-FNRS Belgium for 
financial support.
This research is financially supported by the Belgian Federal Government
(IAP project ``NOSY") and the ``Communaut\'e fran\c caise de Belgique''
(contract ``Actions de Recherche Concert\'ees'' No. 04/09-312).
This work is partially supported by a Grant-in-Aid for Scientific
Research (C) (No.17540365) from the Japan Society of the Promotion of Science,
and by the ``Academic Frontier'' Project at Waseda University
from the Ministry of Education, Culture, Sports, Science and
Technology of Japan.


\end{document}